\documentclass[aps,prb,superscriptaddress,showpacs,twocolumn]{revtex4}
\usepackage{graphicx}
\usepackage{amsmath}

\newcommand{\kB}{k_{\text{B}}}

\begin{document}
\title{$T_c$ for non $s$-wave pairing superconductors correlated with
   coherence length and effective mass}
\author{G. G. N. Angilella} 
\affiliation{Dipartimento di Fisica e Astronomia, Universit\`a degli
   Studi di Catania,\\
and Istituto Nazionale per la Fisica della Materia, Sezione di
   Catania,\\
Corso Italia, 57, I-95129 Catania, Italy}
\author{N. H. March}
\affiliation{Oxford University, Oxford, England}
\affiliation{Department of Physics, University of Antwerp (RUCA), Antwerp,
   Belgium}
\author{R. Pucci}
\affiliation{Dipartimento di Fisica e Astronomia, Universit\`a degli
   Studi di Catania,\\
and Istituto Nazionale per la Fisica della Materia, Sezione di
   Catania,\\
Corso Italia, 57, I-95129 Catania, Italy}

\date{7 July 2000}
\begin{abstract}
For unconventional heavy Fermion superconductors, typified by
   UBe$_{13}$, the superconducting transition temperatures $T_c$ are
   shown to correlate with a characteristic energy $\hbar^2 /(m^*
   \xi^2 )$, $m^*$ being the effective mass and $\xi$ the coherence
   length.
For four of the six materials for which $T_c$, $m^*$ and $\xi$ are
   available, $\kB T_c \sim 20 \hbar^2 / (m^* \xi^2)$.
One heavy Fermion material, UPd$_2$Al$_3$, reveals a tendency for the
   above linear behaviour to saturate at substantially larger $\hbar^2
   / (m^* \xi^2)$ than for UBe$_{13}$.
The sixth material considered, URu$_2$Si$_2$, falls between
   UBe$_{13}$ and UPd$_2$Al$_3$.
To embrace $d$-wave pairing in cuprates, a $\log$-$\log$ plot reveals
   that $\kB T_c \sim \hbar^2 / (m^* \xi^2)$, but more materials for
   which $m^*$ and $\xi$ are measured will be required to substantiate
   the correlation in these high-$T_c$ substances.
\end{abstract}
\maketitle

\begin{figure}
\centering
\includegraphics[height=\columnwidth,angle=-90]{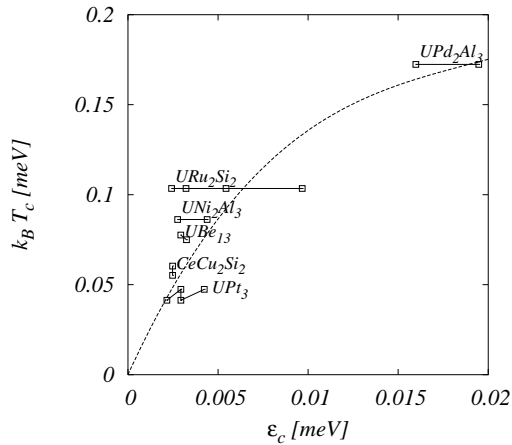}
\caption{Thermal energy $\kB T_c$ corresponding to superconducting
   transition $T_c$ versus characteristic energy $\epsilon_c = \hbar^2
   /(m^* \xi^2 )$, with $m^*$ the effective mass and $\xi$ the
   coherence length (see also Table~\protect\ref{tab:HF}).
Six heavy Fermion materials are considered.
Fitted (dashed) curve is represented in Eq.~(\protect\ref{eq:fit}) and
   should be regarded mainly as a guide to the eye.
}
\label{fig:HF}
\end{figure}

Although heavy Fermion superconductors were discovered more than a
   decade and a half ago\cite{Cox:98,Sauls:94}, interest in their
   physical properties remains considerable.
Thus, in a very recent study\cite{Waelti:00}, the differential
   conductivity of a UBe$_{13}$--Au junction has been measured in both
   superconducting and normal states, yielding in particular an energy
   gap $\Delta$ in this unconventional superconductor very different
   quantitatively from a BCS relation which at $T=0$ reads $2\Delta /
   \kB T_c = 3.5$.
This figure is `enhanced' to around $7$ from this experiment on
   UBe$_{13}$.
We shall return briefly, at the end of this Report, to this matter of
   the energy gap in this non $s$-wave heavy Fermion superconductor.
However, the main focus of the present investigation is to address the
   question as to whether there is a `natural'  energy scale on which
   to measure $\kB T_c$ in non $s$-wave superconductors.
And then, leaving aside $\ell(\ell+1)$ in the eigenvalues of $L^2 /
   \hbar^2$, with $L$ the orbital angular momentum of a
   superconducting Cooper pair, a characteristic
   energy $\epsilon_c$ would appear to be
\begin{equation}
\epsilon_c = \frac{\hbar^2}{m^* \ell_c^2} ,
\label{eq:charener}
\end{equation}
where $m^*$ is the effective mass, while $\ell_c$ is a characteristic
   length that remains to be chosen.
That $m^*$ should enter inversely in determining the scale of $\kB
   T_c$ was clearly recognized in the study of Uemura \emph{et
   al.}\cite{Uemura:91}, who did not, however, address the question of
   the length $\ell_c$ below $T_c$.
In the superconducting state of the heavy Fermion materials which we
   first focus on below, it seemed to us that the natural physical
   choice was to take for $\ell_c$ in Eq.~(\ref{eq:charener}) the
   coherence length $\xi$.
We have then found in the available literature simultaneously data on
   $T_c$, $m^*$, and $\xi$ for the six heavy Fermion systems listed in
   Table~\ref{tab:HF} (see Ref.~\onlinecite{Heffner:95}).
These are the data we have therefore used to construct
   Fig.~\ref{fig:HF}, in which $\kB T_c$ has been plotted against the
   `independent variable' $\hbar^2 / (m^* \xi^2 )$ from
   Eq.~(\ref{eq:charener}) with $\ell_c = \xi$.
That there is a marked correlation between these two energies (both
   measured in meV in Fig.~\ref{fig:HF}) is clear.
The dashed curve, though mainly plotted as a guide to the eye, is
   represented over the range shown by the power series in
   $\epsilon_c$
\begin{equation}
\kB T_c = b \epsilon_c \left( 1 + c_1 \epsilon_c + c_2 \epsilon_c^2
   \right) + {\mathcal O} (\epsilon_c^3 ),  
\label{eq:fit}
\end{equation}
where $b \simeq 22$.
For the three materials shown in Fig.~\ref{fig:HF} with the lowest
   $T_c$ values, $\kB T_c / [\hbar^2 /(m^* \xi^2 )] \simeq 20$, as
   follows from the first term in the fitting series
   Eq.~(\ref{eq:fit}).

Below we shall briefly compare and contrast this linear behaviour at
   low $T_c$ with that for the $d$-wave pairing in the cuprates.
However, the further point to be stressed is that the material
   UPd$_2$Al$_3$, though having somewhat different coherence lengths
   in different crystal directions, shows a clear tendency of the
   (assumed) relation
\begin{equation}
\kB T_c = f_{\text{hF}} \left( \frac{\hbar^2}{m^* \xi^2} \right)
\label{eq:form}
\end{equation}
to go from the linear form $\sim 20 \hbar^2 / (m^* \xi^2 )$ at small
   argument to $f_{\text{hF}} \to \text{const}$ in these heavy Fermion
   (hF) materials as the independent variable $\epsilon_c$ is
   increased by a factor of about $5$ from UBe$_{13}$ to
   UPd$_2$Al$_3$.

In the light of the above findings for the unconventional heavy
   Fermion superconductors considered in Table~\ref{tab:HF} and
   Fig.~\ref{fig:HF}, it seemed of obvious interest to compare and
   contrast their behaviour with corresponding results for the
   high-$T_c$ cuprates, known also to have non $s$-wave pairing.
But then the difficulty comes up that for only very few cuprates are
   data simultaneously available on the same materials for $T_c$,
   $m^*$, and $\xi$, entering the correlation proposed in this Report.

\begin{figure}
\centering
\includegraphics[height=\columnwidth,angle=-90]{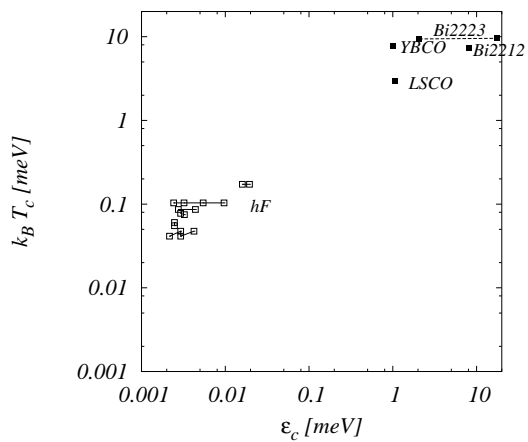}
\caption{$\log$-$\log$ plot of Fig.~\protect\ref{fig:HF}, but now with
   high-$T_c$ cuprates in top-right hand corner (data taken from
   Poole \protect\emph{et al.}\protect\cite{Poole:95}), in addition to
   heavy Fermion data of Fig.~\protect\ref{fig:HF} (here collectively
   marked by `hF').
For the cuprates, use has been made of the coherence length $\xi_{ab}$.
}
\label{fig:loglog}
\end{figure}

Nevertheless, in spite of the sparseness of the data, we felt it of
   obvious interest to show in Fig.~\ref{fig:loglog} a $\log$-$\log$
   plot in which $\kB T_c$ is again displayed versus $\hbar^2 / (m^*
   \xi^2 )$.
The plot is, to our mind, sufficiently encouraging to warrant further
   work in measuring both $m^*$ and $\xi$ on other high-$T_c$
   cuprates.
The striking difference from the heavy Fermion cases is that now,
   accepting the wide spread of data, $\kB T_c \sim \hbar^2 / (m^*
   \xi^2 )$, which is, roughly speaking, one order of magnitude
   different from the linear limit of Eq.~(\ref{eq:fit}) for the heavy
   Fermion materials.

\begin{figure}
\centering
\includegraphics[height=\columnwidth,angle=-90]{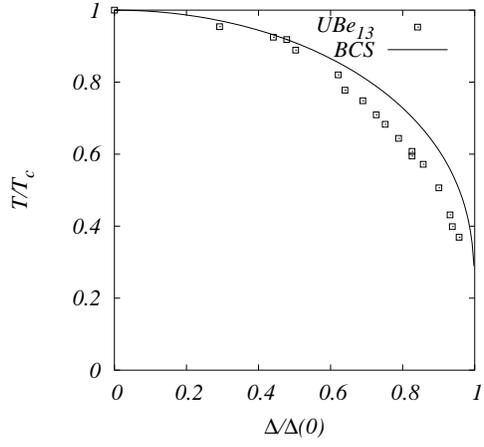}
\caption{$T/T_c$ (on ordinate) vs $\Delta(T)/\Delta(0)$.
Points are experimental data for UBe$_{13}$ (from W\"alti \protect\emph{et
   al.}\protect\cite{Waelti:00}), with $\Delta(0)$ extrapolated to
   $\sim 7 \kB T_c$, while the solid line is BCS curve.
}
\label{fig:UBe13}
\end{figure}

To return briefly to the new measurements of W\"alti \emph{et
   al.}\cite{Waelti:00} on the energy gap $\Delta(T)$ in UBe$_{13}$,
   we have plotted their data in somewhat unorthodox form in
   Fig.~\ref{fig:UBe13}, with $T/T_c$ on the ordinate, and
   $\Delta/\Delta(0)$ on the abscissa.
Furthermore we have, admittedly with some small degree of
   arbitrariness, extrapolated the measured data to pass through twice
   the BCS value.
What we wish to emphasize, in the present context of non $s$-wave
   pairing superconductors, is that the renormalized BCS curve near
   $T=0$ can be referred to as a `gapped' phase, whereas the
   experimental curve shows excitations (gapless as well as gapped)
   characteristic of non $s$-wave pairing.
We expect, near $T=T_c$, that the difference between the BCS and the
   UBe$_{13}$ curves will reflect in general terms the specific heat
   low-temperature behaviour in the normal state of UBe$_{13}$, namely
\begin{equation}
C_V = \gamma T + B T^3 ,
\end{equation}
but it would take us well beyond the scope of the present study to
   attempt further, quantitative analysis on this issue.

In summary, motivated by recent continuing interest\cite{Waelti:00} in
   heavy Fermion materials like UBe$_{13}$ and related
   compounds\cite{Martisovits:00}, we have reopened the question as to
   whether there is a `natural' energy scale on which to measure $\kB
   T_c$.
For the heavy Fermion cases with the lowest transition temperatures,
   we have presented evidence that $\kB T_c \sim 20 \hbar^2 / (m^*
   \xi^2 )$ in these non $s$-wave pairing superconductors.
However, over a wider range of $\hbar^2 / (m^* \xi^2 )$ for these
   materials, the form Eq.~(\ref{eq:form}) has been proposed, where
   $f_{\text{hF}} (x) \to \text{const}$ for values of $x$ substantially
   larger than the value for, say, UBe$_{13}$ ---a situation which
   occurs in fact for UPd$_2$Al$_3$.
Fig.~\ref{fig:loglog} shows the non $s$-wave cuprates on the same
   diagram as the heavy Fermion superconductors, and now, roughly
   speaking, we are in a regime where $\kB T_c$ and $\hbar^2 / (m^*
   \xi^2 )$ are the same to better than order of magnitude.

\begin{acknowledgments}
One of us (N. H. M.) made his contribution to the present Report
   during a visit to the Physics Department, University of Catania, in
   the year 2000.
Thanks are due to the Department for the stimulating environment and
   for much hospitality.
N. H. M. also wishes to thank Professor V. E. Van Doren for his
   continuous interest and support.
G. G. N. A. acknowledges support from the EU through the FSE program.
\end{acknowledgments}

\bibliographystyle{apsrev}
\bibliography{HF}

\begin{table*}
\caption{Selected physical properties for six heavy Fermion materials
   [from Heffner and Norman (Ref.~\protect\onlinecite{Heffner:95})]}
\begin{ruledtabular}
\begin{tabular}{l|rrrrrr}
  & UPt$_3$ & UBe$_{13}$ & UNi$_2$Al$_3$ & UPd$_2$Al$_3$ &
   URu$_2$Si$_2$ & CeCu$_2$Si$_2$ \\
\hline
$T_c$ [K] & 0.55 & 0.9 & 1.0 & 2.0 & 1.2 & 0.7 \\
$\xi$ [\AA] & \begin{tabular}{r} 100 ($\parallel ab$) \\ 120
   ($\parallel c$) \end{tabular} & 100 & 240 &
   85 & \begin{tabular}{r} 100 ($\parallel ab$) \\ 150 ($\parallel c$)
   \end{tabular} & 90 \\
$m^* / m_e$ & 180 & 260 & 48 & 66 & 140 & 380 \\
\hline
\end{tabular}
\end{ruledtabular}
\label{tab:HF}
\end{table*}

\end{document}